# Extreme-ultraviolet laser generation at 118 nm via adaptive random additional periodic-phase engineering in a LiF crystal

Yanling Cheng[1], Bin Zhang[2], Fei Liang[1], Haohai Yu[1,*], Huaijin Zhang[1,*]

[1] State Key Laboratory of Crystal Materials and Institute of Crystal Materials, Shandong University, Jinan 250100, China

[2] State Key Laboratory of Crystal Materials and School of Physics, Shandong University, Jinan 250100, China

*Corresponding author: haohaiyu@sdu.edu.cn, huanjinzhang@sdu.edu.cn

**ABSTRACT:** Extreme ultraviolet (EUV) coherent sources below 120 nm are of paramount significance for promoting next-generation nanoscale lithography, precision spectroscopy, and exploring the emerging physical phenomena in quantum materials. Nonlinear optical conversion serves as the only feasible approach to obtain solid-state EUV lasers, yet the intrinsic strong absorption at EUV and giant phase mismatch among light waves have hindered the realization of highly-efficient EUV light sources. Herein, we propose a random additional periodic phase (RAPP) strategy in third-order nonlinear crystals to overcome these problems, that an artificially-designed random phase grating at micrometer-scales is embedded in the homogeneous bulk crystal, thus adaptively compensating the phase mismatch between fundamental-wave and third-harmonic waves. For the first time, the EUV laser at 118 nm is demonstrated in the RAPP lithium fluoride (LiF) crystals with wide period distributions, where the highest output power is over 90 μW. To the best our knowledge, this is the shortest wavelength among all all-solid-state laser systems, which represents a significant advance in nonlinear optical materials and opens new roadmap toward high-brightness EUV sources.

## Introduction

Lasers have been widely applied in our daily life. Exploring laser sources at new wavelengths has been an enduring pursuit in optical science and engineering, which opens great opportunities for fusion energy, precision spectroscopy, photoacoustic imaging, and so on. At present, the practical laser sources at ultraviolet, visible, and mid-infrared region (300~2000 nm) have been rapidly developed and utilized. Benefitting from the giant single photon energy over 10 eV, extreme ultraviolet (EUV) light below 120 nm would be an indispensable optical tool for semiconductor manufacturing, ultrafast science, electron dynamics and advanced materials characterization. For example, if we have a high-brightness EUV light, the momentum detection range of angle resolved photoelectron spectrometer (ARPES) could cover the first Brillouin zone of the high-temperature superconducting cuprates, thus opening new routes to solve the long-standing Holy Grail of condensed matter physics. However, at present, EUV light below 120 nm, especially all solid-state laser source, is still absent. In order to obtain EUV light, some cutting-edge optical systems were developed, including synchrotron radiation, plasma discharges, and gas high-order harmonic generation (HHG), but suffering from complex setup and high cost.

Nonlinear optical conversion is an effective method to extend the practical laser wavelengths, as demonstrated in ultraviolet BBO, LBO, and KBBF crystals. However, at EUV region below 120 nm, none of these commercialized NLO crystals can be utilized owing to the intrinsic optical absorption. Even in some EUV transparent NLO crystals, *i.e.* $SrB_4O_7$ and $BaMgF_4$, the weak optical anisotropy, corresponding to small birefringence, also hinders the realization of highly-efficient laser frequency conversion. Therefore, the all-solid-state EUV lasers based on nonlinear effects in bulk crystals remains a great challenge despite of its compelling advantages in terms of compactness and integrability.

Herein, we proposed a novel random additional periodic phase (RAPP) engineering to make use of varied periodic lengths, thus creating the adaptive phase compensation between fundamental-wave and high-order harmonic waves. A random factor $X_j(z')$ was introduced into the coupled wave equations. This phase engineering strategy contains more Fourier components, allowing for spatial frequency diversity to satisfy the phase matching conditions of the nonlinear crystals at the edge of the EUV bandgap. At this time, nonlinear frequency conversion can be enhanced at any periods, thus breaking the long-standing challenge of precision machining on NLO crystal materials. Using femtosecond laser direct writing technology, we fabricated three RAPP lithium fluoride (LiF) samples with periodic distributions of 2.5 μm, 3.6 μm and 4.7 μm. For the first time, the EUV laser at 118 nm was realized from three RAPP-LiF crystals, indicating the validity of our random phase engineering strategy. The highest output power is 90 μW, which is several orders of magnitude higher than that of random QPM $SrB_4O_7$ crystal. To our knowledge, this is the shortest wavelength among all all-solid-state laser systems, which would greatly push the developments of EUV equipment, *e.g.* ARPES, nuclear clock, and EUV Raman spectroscopy.

**Results and Discussion**

In the field of nonlinear optics, the third-order nonlinear susceptibility $\chi^{(3)}$ governs a wide range of phenomena, including the Kerr effect, self-phase modulation (SPM), four-wave mixing (FWM), and third-harmonic generation (THG), etc. Unlike the second-order processes that are only generated by non-centrosymmetric materials, the interactions mediated by $\chi^{(3)}$ are universally present in all materials and provide an intrinsic pathway for frequency up-conversion. However, the efficiency of THG in bulk crystal is often limited by the material dispersion and momentum mismatch. In the EUV wavebands, RAPP emerged as a powerful physical concept, achieving output through random forms of spatial phase modulation within certain range. It not only compensates for the accumulated phase mismatch over a long interaction length to realize efficient third-order frequency conversion, but also enables customized spectral control of harmonic generation.

For the THG effect in centrosymmetric medium, the polarization response involves nonlinear effect triggered by the cubic power term of the electric field:

$$P^{(3)}(3\omega, z) = \varepsilon_0 \chi^{(3)}(\omega, \omega, \omega) \vdots E_\omega(z)E_\omega(z)E_\omega(z) \tag{1}$$

where $\varepsilon_0$ denotes the vacuum permittivity, $\chi^{(3)}$ is the third-order susceptibility tensor, $\omega$ refers to the fundamental frequency, and $E_\omega(z)$ denotes the electric field of the fundamental light at the propagation length $z$.

The coupled-wave equation for THG process can be described as:

$$\frac{dE(3\omega, z)}{dz} = i\frac{3\omega}{2cn_{3\omega}}\chi^{(3)}(\omega, \omega, \omega)E^3(\omega)\exp(i\Delta\varphi) \tag{2}$$

where $n_{3\omega}$ refer to the refractive index of THG light, $c$ is the light velocity, and $\Delta\varphi$ represents the intrinsic phase mismatch.

In previous APP strategy applied in second-order nonlinearity, the effective nonlinearity coefficient $d_{eff}(z)$ is arranged in 0/1/0/1/0/1 periodic alternating pattern to engineer reciprocal lattice vectors to compensate phase mismatch

$$\Delta\varphi = \Delta kz = (k_{3\omega} - 3k_\omega)z = 2\pi\left(\frac{n_{3\omega}}{\lambda_{3\omega}} - \frac{3n_\omega}{\lambda_\omega}\right)z \tag{3}$$

However, in the EUV band, the required APP period $\Lambda_{APP} = \frac{2\pi}{\Delta k} = 1 / \left(\frac{n_{3\omega}}{\lambda_{3\omega}} - \frac{3n_\omega}{\lambda_\omega}\right)$ is below one hundred nanometers, beyond the current preparation capability of femtosecond laser direct writing.

RAPP strategy breaks the limitation by interlacing "ordered" (high-$d_{eff}$) and "disordered" (low-$d_{eff}$) regions with random factor "$X_j(z')$" in the crystal materials, not the strict periodic structure. The nonlinearity coefficient of RAPP is modeled as:

$$\chi^{(3)}(z) = \begin{cases} \chi_a^{(3)}, & z \in ordered\ domains \\ \chi_b^{(3)}, & z \in disoredred\ domains \end{cases} \tag{4}$$

The segment lengths of RAPP crystal are drawn from a distribution $\{L_j\}$, and phase perturbation $X_j(z')$ imposed on each segment boundary. Under the small-signal approximation, the accumulated THG amplitude generated in RAPP phase engineering can be expressed as:

$$E(3\omega, z) = \frac{3i\omega}{2cn_{3\omega}} E_\omega^3(z) \left[ \int_0^z \chi_{eff}^{(3)}(z) e^{i\Delta kz} e^{i\Delta k X_j(z')} dz \right]$$

$$= \frac{3i\omega}{2cn_{3\omega}} E_\omega^3(z) \left[ \int_0^{z_1} \chi_a^{(3)} e^{i\Delta kz} e^{i\Delta k X_{j1}} dz + \int_{z_1}^{z_2} \chi_b^{(3)} e^{i\Delta kz} e^{i\Delta k X_{j1}} dz + \int_{z_2}^{z_3} \chi_a^{(3)} e^{i\Delta kz} e^{i\Delta k X_{j2}} dz + \right.$$

$$\left. \int_{z_3}^{z_4} \chi_b^{(3)} e^{i\Delta kz} e^{i\Delta k X_{j2}} dz + \int_{z_4}^{z_5} \chi_a^{(3)} e^{i\Delta kz} e^{i\Delta k X_{j3}} dz + \cdots \cdots \right] \quad (5)$$

where $\chi_a^{(3)}$ and $\chi_b^{(3)}$ represent the effective nonlinear coefficients of the ordered and disordered regions respectively; $\Delta kz$ is the phase difference between the interacting waves, and $\Delta k X_j(z')$ represents a spatially random phase modulation term of the RAPP structure. By introducing the continuous spatial frequencies through the random phases $X_j(z')$, the Fourier transform of the nonlinear coefficient $\tilde{d}(k) = \mathcal{F}\{d(z)e^{i\Delta\varphi}\}$ has non-zero components near $k \approx \Delta k$, thereby adaptively compensating for phase mismatch without requiring sub-wavelength precision.

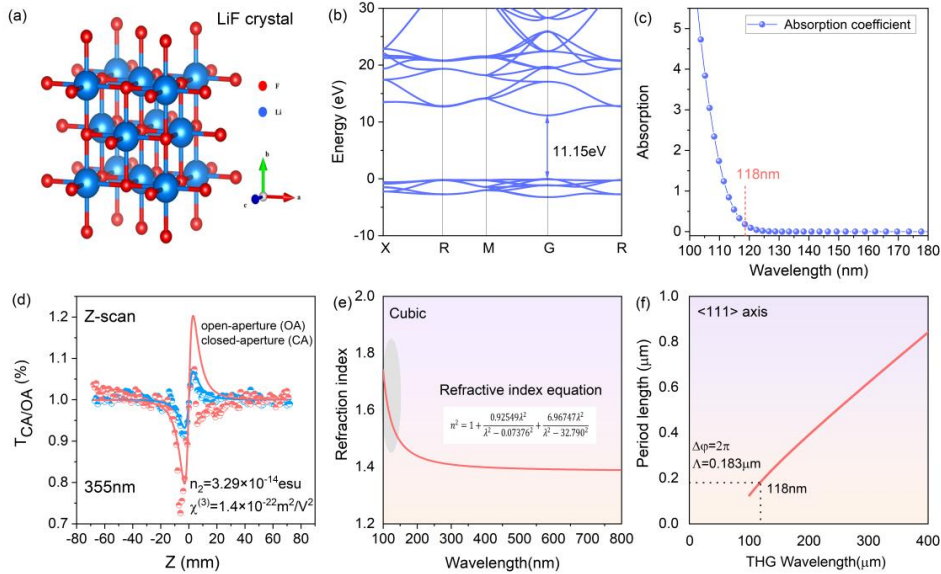

**Figure 1.** (a) Crystal structure of cubic LiF. Li$^+$ ions (blue balls) and F$^-$ ions (red balls) form the six-coordinate structure with octahedral symmetry arrangement. (b) The calculated electronic structure of LiF crystal by HSE06 hybrid functionals. (c) The calculated absorption coefficients of LiF crystal at EUV region around 118 nm. (d) Z-scan measurements at 355 nm for third-order nonlinear optical properties, under open aperture (OA) and closed aperture (CA) conditions. The fitted $\chi^{(3)}$ is 1.4×10$^{-22}$ m$^2$/V$^2$. (e) Refractive index of LiF crystal. (i) The calculated periodic length of RAPP LiF crystal for THG process (355→118 nm). $\Delta\varphi = \Delta kz = 2\pi$, $\Lambda_{\text{LiF}}$=0.183 μm from 355 nm to 118 nm.

**Figure 1** presents the comprehensive physical and optical suitability of lithium fluoride (LiF) as a nonlinear medium for EUV laser. The face-centered cubic LiF lattice (space group Fm$\bar{3}$m), in which Li$^+$ ions and F$^-$ ions are hexatically coordinated in an octahedral symmetric arrangement, forms with the highly ionic Li-F bonds (**Fig.1a**). As shown in **Fig. 1b**, the electronic band structure of LiF crystal indicates that the maximum value of the valence band (VB) occurs at the F point, while the minimum value of the conduction band (CB) is located at the gamma-point. The indirect band gap of LiF is calculated to be 11.15 eV, placing the intrinsic absorption edge below 110 nm. The wavelength-dependent absorption coefficient of the LiF crystal was derived based on first-principles methods, from the frequency-dependent dielectric function $\varepsilon(\omega)$. As despicted in **Fig. 1c**, at the wavelength around 120 nm, the predicted absorption rates were approximately 0.1, which demonstrates the minimal linear loss of LiF crystals at the EUV wavelength range.

The third-order nonlinear optical properties of LiF crystals were systematically characterized using the Z-scan technique at a wavelength of 355 nm. Both open-aperture (OA) and closed-aperture (CA) measurements were performed, combined with data fitting analysis, the nonlinear refractive index of this crystal was obtained as $n_2 \approx 3.29 \times 10^{-16}$ esu, corresponding to the third-order nonlinear susceptibility was calculated as $\chi^{(3)} \approx 1.4 \times 10^{-22}$ m$^2$/V$^{-2}$ (**Fig. 1d**). The Sellmeier dispersion relation for LiF is shown in **Fig. 1e**, it is apparent that near the absorption edge, the refractive index of LiF exhibits distinct steepening characteristic. Therefore, in the EUV region, even minute deviations of wavelength can result in substantial phase-velocity mismatch, which highlights the serious dispersion problem faced by traditional phase matching schemes, Then, we designed the third-harmonic LiF device ranging from 355 nm to 118 nm (**Fig. 1f**). Based on the phase-matching condition, the idealized phase grating period is $\Lambda_{LiF} = 0.183$ μm, far below the minimum fabrication size of femtosecond laser direct writing, so the RAPP strategy is introduced by engineering micron-scale phase functional units within bulk crystal. Taken together, these characteristics validate LiF as an outstanding nonlinear host for EUV generation.

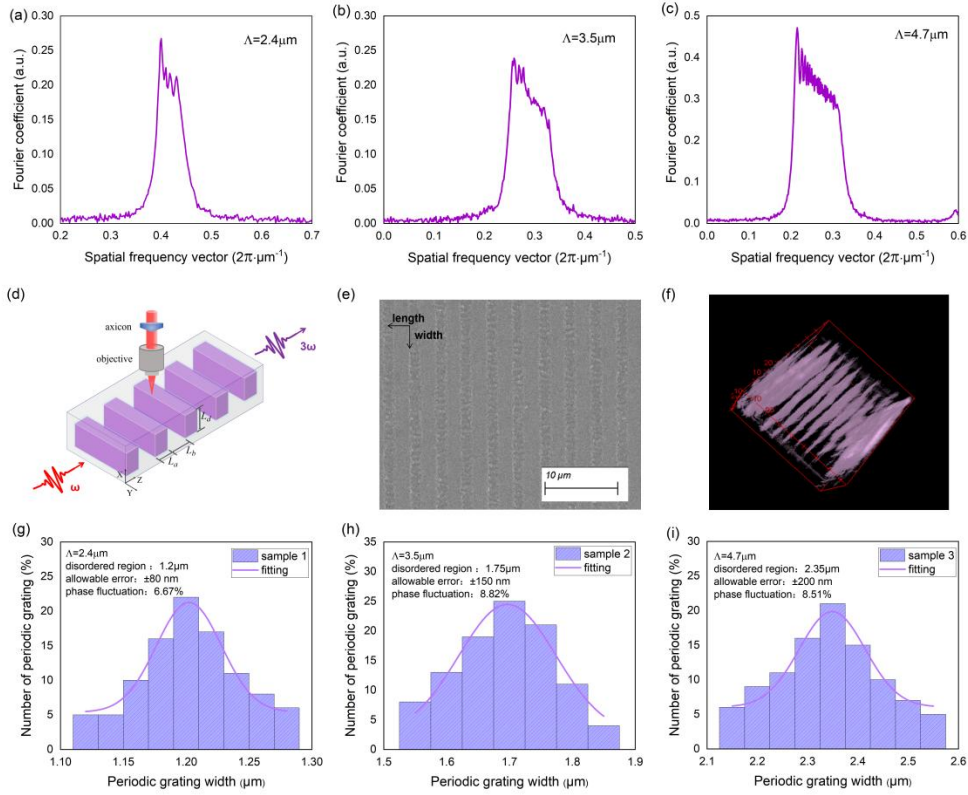

**Figure 2.** (a-c) The spatial frequency components for each RAPP-LiF structural configuration of 2.5 μm, 3.6 μm, and 4.7 μm through Fourier transforms. (d) Schematic diagram of femtosecond laser direct writing process. (e) SEM graph of RAPP LiF crystal. (f) Three-dimensional refractive index distribution in RAPP LiF crystal. (g-i) Periodic lengths distribution histograms in three different RAPP LiF crystal. Λ= 2.5 μm, 3.6 μm, 4.7 μm.

In order to meet the RAPP condition in LiF crystals and take into account the spatial resolution of femtosecond laser direct writing, we systematically designed three different microstructures with phase grating periods, namely 2.4 μm, 3.5 μm and 4.7 μm, respectively. By applying Fourier transforms to each structural configuration, the corresponding spatial frequency spectra are obtained, as depicted in **Figs. 2a-c**. These spectra reveal the spatial frequency components of the effective nonlinear polarization in the RAPP crystals. Notably, multiple random auxiliary frequency components emerge near the primary frequency terms, resulting in a significantly broadened spectral profile in the Fourier domain, which effectively expands the phase-matching bandwidth for THG effect in the LiF crystal.

**Figure 2d** illustrates the RAPP microstructures fabrications through femtosecond laser direct writing technology, in which the femtosecond laser beam is focused on LiF crystal through an axicon and objective lens. This scanning electron microscope (SEM) image in **Fig. 2e** shows the actual grating structure, with a design period of 3.6 μm. The darker areas correspond to the disordered regions for frequency conversion, with a minimum lateral width approximately 1.2 μm. As clearly revealed by the SEM images, the overall grating period remains stable, and the random factors '$X_j(z')$' in the RAPP structure change the periodicity of written region. This phase engineering structure disrupts the previous perfectly periodic arrangement with a duty cycle of 1:1, regulating the effective response of the nonlinear frequency conversion regions within each period, and performing the phase compensation in the disordered areas automatically. To analyze the impact of laser writing on the optical properties of the material, a 633 nm wavelength laser beam was used to pass through the LiF crystal, and then compare the differences in interference fringes between the characterized and non-characterized regions. The 3D refractive index profile of the RAPP LiF sample was shown in **Fig. 2f**, and the refractive index variation within the written regions are within the range of 0.004 to 0.007.

**Figs. 2g-i** display the statistical histograms of written dimensions in LiF sample, designed with single periods of 2.5 μm, 3.6 μm, and 4.7 μm, respectively. When the designed grating period is 2.4 μm, the disordered length are approximately distributed within the range of 1.12 to 1.28 μm, the allowable error of the phase grating is ±80 nm, and the corresponding phase fluctuation is ±6.67%. Similarly, when the designed phase grating period is 3.5 μm and 4.7 μm, the allowable error of the phase grating is ±150 nm and ±200 nm, and the corresponding phase fluctuations are ±8.82% and 8.51%, respectively. These data show the actual distribution of the phase grating periods and the variability of the femtosecond laser direct writing process, providing important references for the precision and repeatability of laser manufacturing.

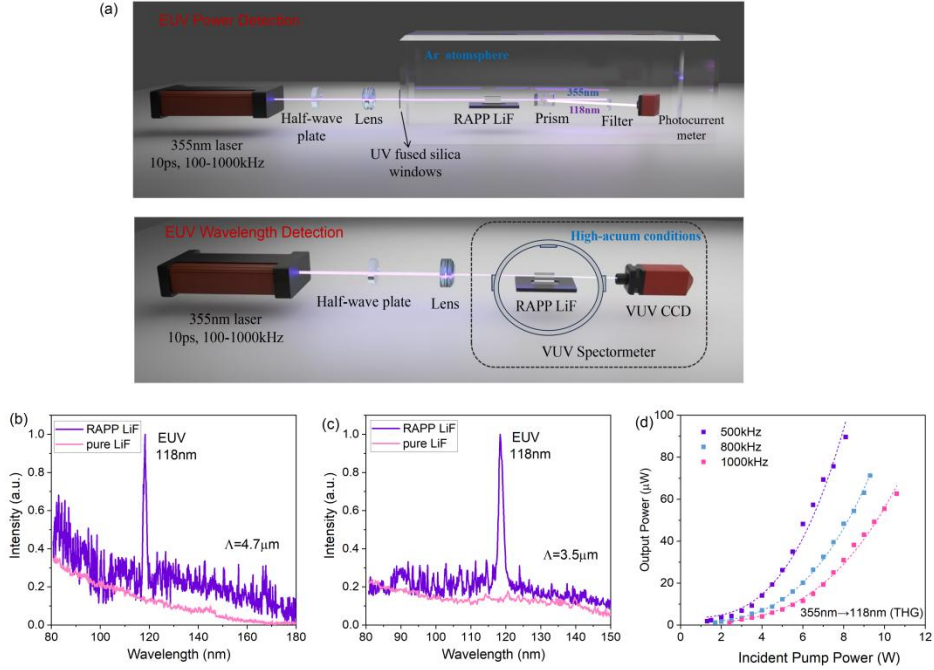

**Figure 3.** (a) Schematic graph for THG laser experiments in RAPP LiF crystal. *Top*: Power measurements, *Bottom*: laser wavelength detection. (b, c) EUV laser spectrum at 118 nm. $\Lambda$=4.7 μm, and $\Lambda$=3.5 μm. It is observed that there is no THG signal in bulk LiF crystal without phase modulation. (d) Output power of EUV laser in RAPP LiF crystal with a design period of 3.5 μm.

**Figure 3a** illustrates the experimental setup for EUV laser experiment. A 355 nm pulse laser was employed as the pump source, with a pulse duration of 10 ps and repetition rate between 100-1000 kHz, to generate 118 nm laser radiation by THG conversion. The incident light is controlled in polarization state by a half-wave plate, and then focused onto the RAPP LiF crystal through a convex lens. Under the experimental conditions, the rayleigh length of the 355 nm beam is 14 mm, which is longer than the length of the LiF sample, and the waist radius is $w_0$=40 μm, smaller than the longitudinal depth $L_d$, to fully confined within the artificially designed grating structures. We use a LiF prism to separate the fundamental light and THG signals and collect the output power of THG signal by a specific photocurrent meter. During the laser experiments, the LiF sample, prism and the power meter are placed inside a sealed box filled with high-purity argon gas, to eliminate absorption losses caused by air in the EUV region. In addition, the THG laser wavelength is recorded by a vacuum ultraviolet spectrometer under vacuum conditions.

**Figs. 3b and 3c** present the normalized EUV laser spectra generated in the RAPP-LiF samples with grating period of 4.7 μm and 3.5 μm, respectively. A characteristic spectral peak at 118 nm is observed in both cases. To the best our knowledge, this is the first solid-state EUV laser below 120 nm realized by nonlinear optical frequency technology. Compared with bulk LiF crystal without gratings, the RAPP samples exhibit a substantial enhancement in THG signal, thereby demonstrating the validity of RAPP strategy in boosting nonlinear conversion efficiency for EUV laser generation. Notably, the sample with Λ=3.5 µm grating period produces stronger THG output than that of sample with Λ=4.7 µm period. This difference is attributed to the phase-matching conditions $\Delta\varphi=2N\pi$ governed by the effective modulation period. The conversion efficiency is inverse proportional to the square of the order $N$, namely $\eta \propto \frac{1}{N^2}$, so the smaller the value of $N$ (shorter period length), the higher the nonlinear conversion efficiency $\eta$. However, in RAPP-LiF sample with Λ=2.4 µm, there is no enhanced THG signal observed, possibly owing to large optical loss at the order/disorder interfaces in short-period sample.

**Figure 3d** depicts the THG output power under different repetition frequencies of the RAPP LiF crystals. It can be observed that, with the same average pump power, the nonlinear conversion efficiency at low repetition frequency is significantly higher than that of high repetition, which could be assigned to the high single pulse energy at the low repetitions. When the incident pump power is 8.1 W and the repetition frequency is set to 500 kHz, the maximum output power at 118 nm reaches up 90 µW, with an overall efficiency of $1.1\times10^{-5}$.

**Conclusion**

In summary, we present the first EUV solid state laser generation at 118 nm via a RAPP engineering in lithium fluoride crystal. Benefiting from the adaptive phase compensation capacity, the long-standing challenge of phase-mismatch in EUV nonlinear optics was overcome, thus giving an effective EUV laser output over 90 µW. More importantly, this technology could be extended to other EUV-transparent oxide and fluoride crystals, *i.e.* $CaF_2$, $BaMgF_4$, and $B_2O_3$, thereby opening great opportunities for EUV laser sources generation and applications. More importantly, using these EUV sources, we can develop new EUV precise instruments for momentum detection, time-space resolved spectrometer, and exploration of those exotic quantum materials, e.g. high-temperature superconductors, topological insulators, and alter-magnetism crystals.

**Methods**
**1. RAPP-LiF sample fabrication**

High-purity LiF crystals were cut and optically polished along the (100), (110), and (111) directions, with dimension of 3×2.8×3.5 mm³ and 3×2.8×5 mm³. Using a Yb femtosecond laser (FemtoYL™-25) with central wavelength of 1030 nm, repetition frequency of 25 kHz, and pulse width of 200 fs, RAPP structures were fabricated inside the crystals through femtosecond laser direct writing technology. The 1030 nm laser beam was focused into the samples through a beam-conditioning system, comprising half-wave plate, aperture, a set of focusing lenses, mirrors, and high-numerical-aperture objective lens. The samples were put onto a PC-controlled 3D XYZ electric-moving stage, moving stage is set as 10 mm/s during laser writing. Three types of grating structures with grating periods of 2.4 μm, 3.6 μm and 4.7 μm, each consisting of 800 periods, were designed. The depth of single layer structure was 50 μm, and the layer spacing was 0.5 mm.

## 2. Sample characterization

### 2.1 Third-order nonlinear optical properties by Z-scan technique

The Z-scanning technique based on the principle of spatial beam distortion was chosen to measure the nonlinear absorption (NLA) characteristics and nonlinear refractive index of the LiF crystal. The excitation laser used was a picosecond (ps) laser with a wavelength of 355 nm, with pulse width of 20 ps and repetition rate of 10 Hz. The single-pulse energy of the incident laser is 2.98 μJ and 9.24 μJ at the position $z=0$. Simultaneously, the reflectivity takes a symmetrical distribution with respect to the position $z=0$.

### 2.2 3D refractive-index imaging

An advanced in-situ 3D refractive-index imaging system (Innofocus, nanoLAB Holoview 3D-Ri) was used to characterize the refractive index changes at 633 nm in laser-induced tracks and adjacent areas in RAPP LiF crystals. In refractive-index imaging system, a 633 nm He-Ne laser is first divided into two coherent light, one light directly enters into a CCD camera. The other light is focused by a 20× microscopic objective into LiF crystals sample and collected by a 50× oil immersion objective at the other side of sample onto the same CCD camera. This incident light of 633 nm enters into LiF sample along its $z$ direction. By analyzing interference fringes of two coherent light collected by the CCD camera, refractive index changes in femtosecond laser-induced tracks and the surrounding regions will be obtained.